*Original Article*

# An Integrated Framework for DevSecOps Adoption

Akanksha Gupta

*Engineering Manager| Fellow BCS | Fellow RSA| IADAS Associate*

**Abstract -** *Introduction of DevOps into the software development life cycle represents a cultural shift in the IT culture, amalgamating development and operations to improve delivery speed in a rapid and maintainable manner. At the same time, security threats and breaches are expected to grow as more enterprises move to new agile frameworks for rapid product delivery. Meanwhile, DevSecOps is a mindset change that revolutionizes software development by embedding security at each step of the software cycle, leading to resilient software. This paper discusses a framework organization can use to embed DevSecOps swiftly and efficiently into the general IT culture.*

## 1. Introduction

DevOps is defined as the intersection of Development and Operations and is a major cultural change in the tech industry. This marriage has created great strides towards a mutual trust relationship between the two. No industry can change overnight but being aware and adopting DevOps early is an important skill set that goes a long way. DevOps is understanding the software and learning to code while learning to operate and maintain that code simultaneously. It is crucial to remember that all it takes is one security breach for the customers to lose trust in any enterprise. Hence, it is important to focus on maintaining high-security standards.

According to Gartner's research- "It is estimated that at least 95% of cloud security failures through 2022 will be the fault of the enterprise,". Therefore, while developing any application, the developer must not have loose ends that may make an enterprise vulnerable to such attacks. With the recent trends of increasing security threats and breaches, an enterprise must be vested in the security of applications from the start of the development cycle. Gartner predicts that "30% of Critical Infrastructure Organizations Will Experience a Security Breach by 2025" [1]. It is what has led to the introduction of DevSecOps in a lot of enterprises. Gartner defines DevSecOps as - "the integration of security into emerging agile IT and DevOps development as seamlessly and as transparently as possible. Ideally, this is done without reducing the agility or speed of developers or requiring them to leave their development toolchain environment."

DevSecOps is a way of baking in security as part of application development and operations, along with bringing teams together to collaborate and use the power of automation and tooling to build robust and secure applications. With DevSecOps, the development process needs to look at security proactively rather than reactively. The development cycle will embed security testing and bug fixes to expose security vulnerabilities early in the software lifecycle. It has paved the way for innovation, efficient developer velocity, and rapid release cycles while keeping security a top priority. It has helped with faster development, rapid release of features, and the power of agile practices.

An imperative question here is whether security lapses only occur because of technical failures. In a recent Gartner study of 367 IT and business leaders, it was found that 50% of the issues involved people, followed by process (37%), technology (8%), and information (5%). Therefore, it is important to look at DevSecOps from a holistic perspective instead of focusing on a narrow area.

Gartner mentions, "DevOps emphasizes people (and culture) and seeks to improve collaboration between operations and development teams." Microsoft's Donovan Brown defines it as: "The union of people, processes, and products to enable continuous delivery of value to our end users." Hence, it is very important to recruit the right set of individuals who will help champion DevSecOps adoption.

Although enterprises have made significant strides in adopting new business practices like moving to cloud providers, using agile frameworks, etc., organizationally, DevSecOps is often given the least priority in the minds of stakeholders. DevSecOps initiatives often don't have a framework to follow; especially those executives can buy. Gartner predicts that through 2022, 75% of DevOps initiatives will fail to meet expectations due to organizational learning and change issues.

Several large companies such as Amazon, Google, and Netflix have adopted DevSecOps to create a high-performing organization and a platform for continuous improvement.

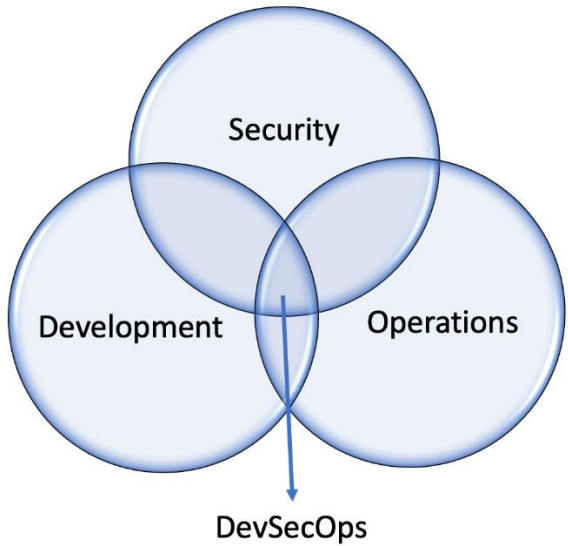

**Fig. 1 DevSecOps Concept**

### *1.1. DevOps vs. DevSecOps*

Both in the core share the idea of a mindset change to focus on development and operations. DevSecOps is an extension of DevOps to bring in the aspect of security early in the development cycle and plan for security mishaps once the deployment is done to production. DevSecOps not only deals with creating a maintainable & operational infrastructure but also addresses the fact that security vulnerabilities are inevitable if enough focus is not paid to security in the early stages of software development.

## 2. DevSecOps Adoption Framework

Enterprise adoption of DevSecOps is a long multi-year journey, and starting early on will help the organization in the long run. There is a lot of material available to bring awareness about DevOps and why one should adopt it. Still, little information is available that talks about DevSecOps and a cohesive framework that organizations could use to adopt DevSecOps seamlessly. This paper focuses on covering this gap and sharing a DevSecOps adoption framework.

### *2.1. Understanding the need for DevSecOps*

The first step in the DevSecOps journey is understanding what DevSecOps is and why we need it fully. Once the understanding of DecSecOps is established, the focus should be on evaluating who and how will benefit from the adoption. It will involve a solid assessment of the business use case, resources, and the current pain points of the organization. In this phase, be transparent with tech debt, defects, and bugs. It will help identify improvements and opportunities to determine the defects' root cause. It will help identify the current applications and processes gaps and help evaluate the opportunities. For example, an opportunity to move a manual job of filtering a few parameters from an excel to automating it to run a script for you will save a lot of time and will not be prone to human error.

Key questions one should ask to evaluate this are:
1. What is DevSecOps?
2. How will the organization benefit from DevSecOps, given the time and resources to spend on it?
3. What are the major pain points we are trying to address as part of this change?
4. What will be our success criteria to evaluate our progress?
5. Which applications will be part of the scope of this adoption?
6. What timelines would we like to consider as we go through this journey?
7. How will it impact our other security initiatives?
8. How do we identify the goals to establish a roadmap to adopt DevSecOps?

### *2.2. Recruit champions/ambassadors of DevSecOps*

The next step is to identify a set of ambassadors or champions who are aligned and believe in the mission of DevSecOps in the organization. It is an opportunity to bring passionate people to volunteer for the program who can bring fresh ideas to the table. One should use several channels to promote the opportunity in the organization so you can have a team of people from diverse backgrounds like engineers, operations, security specialists, testers, and managers. One should look for individuals who are motivated, passionate about learning, can thrive in ambiguity, and are good team players. These individuals should feel empowered to communicate and collaborate on a continued basis. They should fully understand their role and responsibilities in the DevSecOps journey. This cohort will help bring the mission to life and advocate for the cause to help with a broader DevSecOps adoption in the enterprise. In this phase, we provide training (in-house or through external vendors) to this cohort to help upskill them and prepare them for the journey ahead.

Few key questions to ask as you embark on the journey to recruit these individuals are:
1. What are the key duties and responsibilities of the champions or ambassadors of DevSecOps?
2. How many champions should we recruit first, to begin?
3. Are there other technical skill sets we need to look for as we recruit the champions?
4. Will this group work together and collaborate on the initiative?

*2.3. Develop a DevSecOps Strategy*

Developing a DevSecOps strategy is the core of bringing an organization-wide change with DevSecOps. Every individual involved will keep a 'security-first' mindset and involve security best practices from the get-go. As you develop the strategy, keep in mind priority, the cost of time and resources, and ensure the effort is time-boxed for successful implementation. The strategy will help bring alignment regarding what needs to be accomplished as part of the adoption.

Key tenets to keep in mind as you develop the DevSecOps strategy:
1. 'Security-first' mindset - Each person involved will be responsible for keeping security as one of the top priorities. Hence, as part of the software development lifecycle, developers will be accountable for security during development and bug fixes of security vulnerabilities. Similarly, the operations team will have run books to combat security threats once the applications are deployed to production.
2. Automation - As part of a strategy, focus on identifying opportunities where manual tasks can be automated to save time and ensure our systems are less prone to human errors.
3. Continuous integration, continuous deployment (CI-CD)- CI-CD is the pinnacle of DevOps, where the systems around your application are strong enough to test the code changes for you and push them to production without any manual intervention. DevSecOps helps eliminate long testing cycles and leads to shorter iterations with faster releases; early feedback in the software development lifecycle is a key underpinning factor to help us get to the appropriate path or direction as early as possible.

Now that we looked at the key tenets of such a strategy, some key questions your strategy should answer are:
1. How should your organization start implementing DevSecOps?
2. What channels will be created to bring awareness to the organization?
3. What are the short-term and long-term plans to bring adoption?
4. How will success be measured?

*2.4. Communicate to leadership to get buy-in*

Leadership and executives will play a critical role in promoting and adopting DevSecOps. Hence, it is of utmost importance to get a buy-in from them to ensure no other business Objective-Key-Result (OKR) impedes the adoption of DevSecOps in the organization. It is where you will share the strategy with the leadership and make them aware of the initial setup costs in terms of time, money, and resources. At the same time, this is an opportunity to share about long-term rewards and their impact on the organization.

Key questions that should be answered in this stage are:
1. Is the leadership aligned with the strategy and the time commitment?
2. How soon can we get the plan to action?

*2.5 Execution*

The core of the work begins in the execution phase. Time is of the essence; hence it is important to start small, collect feedback, and iterate. In this phase, we evaluate the tools available to help speed up the adoption. Some tools that can help kickstart this process are:
- OWASP
- SonarQube
- Fortify
- HashiCorp Vault

Since each organization is different, use this phase to create a proof of concept (POCs) to see what works best for your organization. It will involve a thorough assessment of solutions for the pain points identified above.

A few key questions that should be answered at this stage are:
1. Instead of doing this as a one-off item, how do you integrate DevSecOps tools into the existing developer infrastructure?
2. What guardrails can be set up to ensure developers cannot bypass these systems?
3. How do you ensure that product delivery speed is not hampered by introducing these systems?

*2.6 Audit and measure success*

Feedback is a critical aspect of human life. As a child, we grow up with continuous feedback from our parents, which is instrumental in learning life skills. Similarly, we should ensure constant feedback in the DevSecOps environment for continual improvement. Alarms, Dashboards and monitoring through alerts are a good way to audit the applications and be proactive in scenarios if something goes wrong. It is a good chance to do a retrospective with the team to see what we did well and what can be improved as a team. At the same time, establish an ongoing feedback mechanism such as quarterly or channels where employees can reach out to provide feedback. It can be done in the form of Agile retro or surveys for individuals to fill in. We should ensure governance and guardrails are in place as we measure the success of the adoption of DevSecOps. As we measure success, ensure you don't over-engineer measurements and instead focus on being result-oriented.

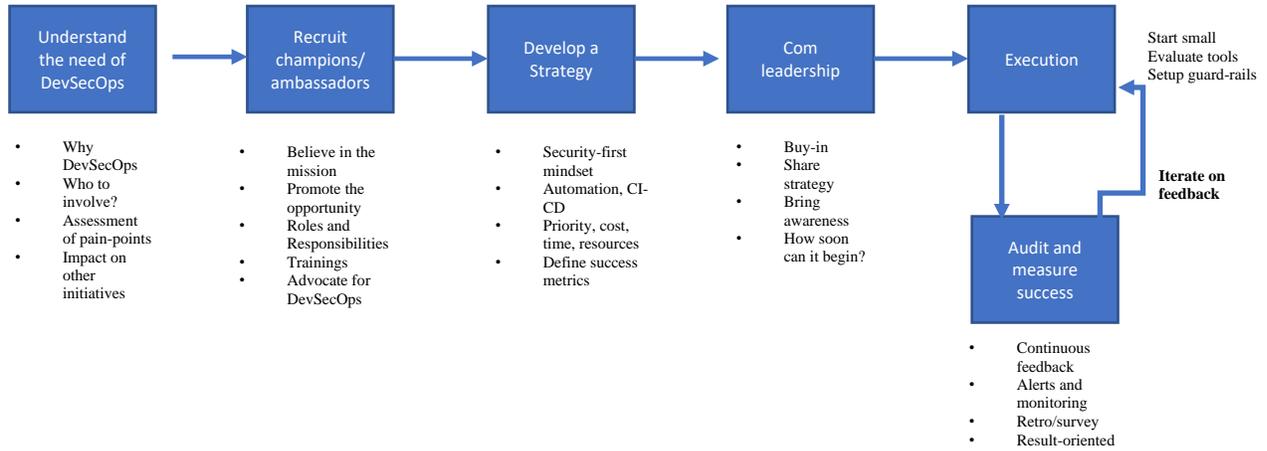

**Fig. 2 DevSecOps Adoption Framework**

A few key metrics to keep in mind are:
- Response times and performance of applications
- Mean time to plan, deploy and deliver
- MTTA (mean time to acknowledge) in times of high severity event
- MTTR (mean time to resolution) in times of high severity event
- Quality of audit reports around open bugs, trouble tickets, deployment reports
- Increased deployment frequency

Based on the feedback gathered in this phase, iterate on the feedback, and go back to execute.

## 3. Conclusion

The framework shared above provides a structured and organized approach that enterprises can use to adopt DevSecOps practices. It helps guide an organization to succeed in its journey to adopt DevSecOps. It will help build secure software from its inception. Investments made in the short term will help the organization with long-term goals to ship better, faster, and more secure products to the customers. Feedback mechanisms will help with continual improvement to iterate and improve. By using this framework, organizations can ensure that common pitfalls for DevSecOps are avoided and ensure that introduction of DevSecOps is a cultural shift in the development process instead of a one-time work.

## Conflicts of Interest

The author(s) declare(s) there is no conflict of interest regarding the publication of this paper.